\documentclass[preprint,12pt]{elsarticle}
\usepackage{graphicx}
\usepackage{amssymb}
\usepackage{amsmath}
\usepackage{url}

\journal{Chaos, Solitons \& Fractals}

\begin{document}

\begin{frontmatter}

\title{Nonlinear Signal Transmission and Harmonic Generation in Plasma Oscillator Chain with Coupling Asymmetry}

\author[srm]{K. Abirami}
\ead{nk.abirami@gmail.com}

\author[urjc]{Miguel A. Prado-Reynoso}
\ead{miguelangel.prado@urjc.es}

\author[urjc,rac]{Miguel A. F. Sanju\'an\corref{cor1}}
\ead{miguel.sanjuan@urjc.es}
\cortext[cor1]{Corresponding author}

\affiliation[srm]{organization={Centre for Nonlinear and Complex Networks, SRM Institute of Science and Technology},
            addressline={Ramapuram},
            city={Chennai},
            postcode={600089},
            state={Tamil Nadu},
            country={India}}

\affiliation[urjc]{organization={Nonlinear Dynamics, Chaos and Complex Systems Group, Departamento de F\'{\i}sica, Universidad Rey Juan Carlos},
            addressline={Tulip\'an s/n},
            city={M\'ostoles, Madrid},
            postcode={28933},
            country={Spain}}

\affiliation[rac]{organization={Royal Academy of Sciences of Spain},
            addressline={Valverde 22},
            city={Madrid},
            postcode={28004},
            country={Spain}}

\begin{abstract} We investigate signal transmission in a one-dimensional chain of $n$ locally coupled plasma oscillators with asymmetric bidirectional coupling $(\epsilon_1,\epsilon_2)$. Our results reveal a clear contrast between symmetric and asymmetric interactions: while symmetric coupling suppresses signal propagation, coupling asymmetry enables efficient nonlinear transmission across the network. In particular, sustained long range transmission is achieved when the forward coupling strength exceeds the backward coupling strength ($\epsilon_1 > \epsilon_2$), as demonstrated through a comprehensive phase diagram in the $(\epsilon_1,\epsilon_2)$ parameter space. Furthermore, under asymmetric diffusive coupling, signal transmission can persist even when the first few oscillators in the chain undergo oscillation death. A central finding of this study is the intrinsic generation of higher harmonics induced by nonlinear interactions, leading to a pronounced reversal of vibrational resonance. This nonlinear redistribution of energy facilitates selective amplification and robust propagation of specific frequency components.
\end{abstract}


\end{frontmatter}

\noindent{\footnotesize \copyright\ 2026. This manuscript version is made available under the CC-BY-NC-ND 4.0 license \url{https://creativecommons.org/licenses/by-nc-nd/4.0/}}

\section{Introduction}
\noindent
Plasma oscillator networks provide a versatile framework for investigating nonlinear dynamics, energy transport, and wave localization in strongly coupled media. A key factor governing these processes is the nature of the coupling between oscillators. In particular, bidirectional coupling is physically relevant, as it allows mutual interaction and energy exchange between neighboring oscillators \cite{karunanidhi2025asymmetric}, in contrast to unidirectional coupling, where energy flow is constrained. Unidirectionally coupled plasma oscillators have been studied in \cite{abirami2026energy}, where fundamental aspects of directed energy transport were reported.

Such interactions are especially important in strongly coupled plasmas, including dusty plasma systems, where charged particles can form ordered chains or lattice structures \cite{dharodi2023ring,lampe2000interactions,merlino2021dusty}. In these systems, long-range Coulomb interactions, modified by Debye shielding, give rise to collective phenomena such as dust-lattice oscillations and nonlinear mode coupling. Similarly, in plasmas supporting ion-acoustic or Langmuir waves, bidirectional interactions are essential for capturing wave reflection, interference, and energy redistribution \cite{wang2022ion}.

In realistic plasma environments, however, coupling is rarely perfectly symmetric. Asymmetry naturally arises due to external fields, spatial inhomogeneities, or sheath effects, leading to nonreciprocal and direction-dependent energy transport. This can induce phenomena such as wave rectification, asymmetric excitation thresholds, and localized modes. Even weak asymmetry may significantly alter system behavior by shifting resonance conditions and destabilizing uniform states \cite{lutin2020nonreciprocal,bhattacharjee2023asymmetric,chen2023nonreciprocity}.

The influence of coupling asymmetry on transport properties has been widely demonstrated across different physical systems. It enables directional heat flow in Floquet systems \cite{74rq-f642}, modifies localization in disordered media \cite{arora2023suppression}, and enhances wave control in mechanical oscillator arrays \cite{kim2023nonreciprocal,kang2022coupling}. In nonlinear lattices and oscillator models, asymmetry gives rise to irreversible transport, localization, and complex collective dynamics \cite{bai2016unidirectional,palacios2021synchronization,search4_kuramoto_asymmetry}. In plasma-related contexts, it has been associated with nonreciprocal energy flow and direction-dependent dynamics \cite{ignatov2025energy,zheng2025thermoacoustic}. These studies establish coupling asymmetry as a key mechanism for tuning nonlinear transport and localization. In particular, it supports nonreciprocal propagation, supratransmission, and breather excitations in various systems \cite{asymmetric_LC,miroshnichenko2011nonlinearly,aguilera2024nonlinear,geniet2002energy,wu2017supratransmission}.

Motivated by these considerations, we investigate signal transmission in a one-dimensional chain of plasma oscillators with asymmetric bidirectional coupling, characterized by distinct forward and backward strengths $\epsilon_1$ and $\epsilon_2$. By systematically exploring the $(\epsilon_1,\epsilon_2)$ parameter space, we identify regimes of efficient signal propagation and attenuation. The results demonstrate that symmetric coupling suppresses transmission, whereas asymmetric coupling enables nonlinear transport accompanied by intrinsic harmonic generation. Furthermore, a reversal of vibrational resonance is observed under specific asymmetric conditions, highlighting the role of nonreciprocal interactions in redistributing energy across frequency modes.

The manuscript is organized as follows. In Sec.~II, we study plasma oscillator networks with symmetric bidirectional coupling, beginning with two coupled oscillators and then extending the analysis to a larger network. In Sec.~III, we address the case of asymmetric coupling strengths, first for two coupled oscillators and then for oscillator chains, where different transmission regimes are identified. In Sec.~IV, we examine higher harmonic generation and the reversal of vibrational resonance. In Sec.~V, we present the transmission phase diagram in the $(\epsilon_1,\epsilon_2)$ parameter space. Finally, the conclusions are given in Sec.~VI.

\section{Plasma Oscillator Networks with Symmetric Coupling}
\subsection{Two Coupled Plasma Oscillators with Symmetric Coupling}

We begin by analyzing a system of bidirectionally coupled plasma oscillators with symmetric coupling. In particular, we consider two identical plasma oscillators interacting through a symmetric bidirectional coupling of strength $\epsilon$. Oscillator~1 is additionally driven by a weak external periodic forcing. The governing equations of motion are given by \cite{kadji2008regular, kadji2012passive, jayaprakash2021doubly, samikkannu2023studies, abirami2026energy}:
\begin{eqnarray}
\ddot{x}_1 + d(1+x_1^2)\dot{x}_1 + \omega_0^2 x_1 + \alpha x_1^2 + \beta x_1^3 &=& F \cos(\omega t) + \epsilon (x_2 - x_1), \nonumber \\
\ddot{x}_2 + d(1+x_2^2)\dot{x}_2 + \omega_0^2 x_2 + \alpha x_2^2 + \beta x_2^3 &=& \epsilon (x_1 - x_2).
\label{eq1}
\end{eqnarray}
The system consists of two identical plasma oscillators characterized by the potential parameters $\omega_0^2$, $\alpha$, and $\beta$, and a nonlinear damping coefficient $d$. The first oscillator, described by the displacement $x_1$, is driven by a weak periodic forcing of amplitude $F$ and frequency $\omega$, while the second oscillator, described by $x_2$, remains at rest in the absence of coupling. The two oscillators are then symmetrically coupled with strength $\epsilon$. Our objective is to analyze the steady-state response of both oscillators under this weak periodic forcing. Accordingly, we seek solutions of the form
\begin{eqnarray}
x_1 &=& A_1 \cos(\omega t + \phi_1), \nonumber \\
x_2 &=& A_2 \cos(\omega t + \phi_2).
\label{eq2}
\end{eqnarray}
Here, $A_i$ and $\phi_i$ ($i=1,2$) denote the amplitude and phase of the steady-state response, respectively. Substituting the ansatz~\eqref{eq2} into Eq.~\eqref{eq1} and applying the method of harmonic balance (back-substitution) \cite{rajasekar2016nonlinear, rajamani2013signal, abirami2026energy}, we obtain the following set of coupled nonlinear algebraic equations for $A_1$, $A_2$, $\phi_1$, and $\phi_2$:
\begin{eqnarray}
A_1^2\left[u_1^2+\left(1+\frac{A_1^2}{2}\right)^2 d^2\omega^2\right]
+ A_2^2\left[\epsilon^2+2\left(1+\frac{A_2^2}{2}\right)^2 d^2\omega^2\right]
- 2 u_1 u_2 A_2^2 &=& F^2, \nonumber \\
A_2^2\left[u_2^2+\left(1+\frac{A_2^2}{2}\right)^2 d^2\omega^2\right]
- \epsilon^2 A_1^2 &=& 0.
\label{eq3}
\end{eqnarray}
with
\begin{eqnarray}
\phi_1 &=& \arctan\left[\frac{\left(1+\frac{A_1^2}{2}\right)(A_1^2+A_2^2)d\omega}{A_1^2u_1-A_2^2u_2}\right], \nonumber \\
\phi_2 &=& \phi_1 - \arctan\left[-\,\frac{\left(1+\frac{A_2^2}{2}\right)d\omega}{u_2}\right],
\label{eq4}
\end{eqnarray}
where
\begin{eqnarray}
u_1 &=& \omega_0^2-\omega^2+\frac{3}{4}\beta A_1^2, \nonumber \\
u_2 &=& \omega_0^2-\omega^2+\frac{3}{4}\beta A_2^2.
\label{eq5}
\end{eqnarray}


The analytical amplitudes $A_1$ and $A_2$ are obtained by solving the nonlinear algebraic system~\eqref{eq3} using the Newton--Raphson method. For the numerical analysis, the response amplitudes are computed as
\begin{equation}
A_i = \sqrt{(Q_i^s)^2 + (Q_i^c)^2},
\label{eq6}
\end{equation}
where
\begin{subequations}
\label{eq7}
\begin{equation}
Q_i^s = \frac{2}{NT} \int_{T_0}^{T_0 + NT} x_i(t)\sin(\omega t)\,dt,
\end{equation}
\begin{equation}
Q_i^c = \frac{2}{NT} \int_{T_0}^{T_0 + NT} x_i(t)\cos(\omega t)\,dt.
\end{equation}
\end{subequations}
Here, $i=1,2$ and $T=2\pi/\omega$ denotes the period of the external forcing. The quantities $Q_i^s$ and $Q_i^c$ are the normalized Fourier sine and cosine coefficients of the steady-state response, evaluated over $N=200$ forcing periods in order to extract the response amplitude at the driving frequency $\omega$. To suppress the influence of transient dynamics, an initial transient interval of duration $T_0=1000$ is discarded prior to the analysis.

\begin{figure}[h]
\begin{center}
\includegraphics[width=0.4\textwidth,height=0.4\textwidth]{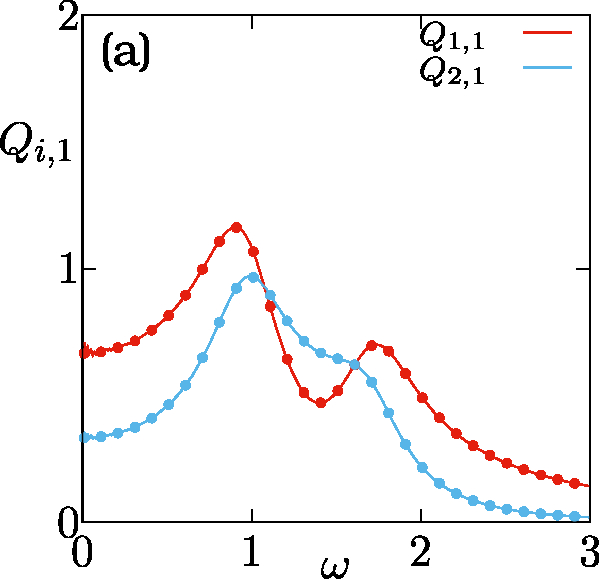}
\end{center}
\caption{Frequency response of two coupled plasma oscillators with symmetric coupling. Solid lines correspond to the numerically computed amplitudes $Q_1$ and $Q_2$, while dots indicate the corresponding analytical predictions. The parameters are $\omega_0^2 = \alpha = \beta = 1$, $d = 0.5$, $\epsilon = 0.5$, and $F = 0.05$.}
\label{fig1}
\end{figure}

The frequency responses of the coupled plasma oscillators, defined as $Q_1 = A_1/F$ and $Q_2 = A_2/F$, are shown in Fig.~\ref{fig1}. The potential parameters are fixed at $\omega_0^2 = \alpha = \beta = 1$, and the nonlinear damping coefficient is set to $d = 0.5$. The driving frequency is varied to obtain the frequency response, while the forcing amplitude is fixed at $F = 0.05$ to ensure a weak-input regime. The symmetric coupling strength is taken as $\epsilon = 0.5$.

In Fig.~\ref{fig1}, the solid lines represent the numerically computed amplitudes, whereas the analytical predictions are indicated by dots. The close agreement between the two demonstrates the consistency between numerical simulations and analytical results over the frequency range considered. For this set of parameters, both oscillators exhibit a nonlinear resonance response. It is also observed that $Q_2$ follows a trend similar to that of $Q_1$, but with comparatively smaller amplitudes.

In the following, we extend the above analysis to a network of bidirectionally coupled plasma oscillators.

\subsection{Network of Bidirectionally Coupled Plasma Oscillators with Uniform Coupling}

The equations of motion for a network of $n$ bidirectionally coupled plasma oscillators, in which only the first oscillator is driven by a weak periodic forcing and all oscillators are coupled with uniform strength $\epsilon$, are given by
\begin{eqnarray}
\ddot{x}_1 + d(1+x_1^2)\dot{x}_1 + \omega_0^2 x_1 + \alpha x_1^2 + \beta x_1^3 &=& F \cos(\omega t) + \epsilon (x_2 - x_1), \nonumber \\
\ddot{x}_i + d(1+x_i^2)\dot{x}_i + \omega_0^2 x_i + \alpha x_i^2 + \beta x_i^3 &=& \epsilon (x_{i-1} + x_{i+1} - 2x_i), \nonumber \\
\ddot{x}_n + d(1+x_n^2)\dot{x}_n + \omega_0^2 x_n + \alpha x_n^2 + \beta x_n^3 &=& \epsilon (x_{n-1} - x_n),
\label{eq8}
\end{eqnarray}
for $i = 2,3,\dots,n-1$.

\begin{figure}[h]
\begin{center}
\includegraphics[height=0.405\textwidth,width=0.4\textwidth]{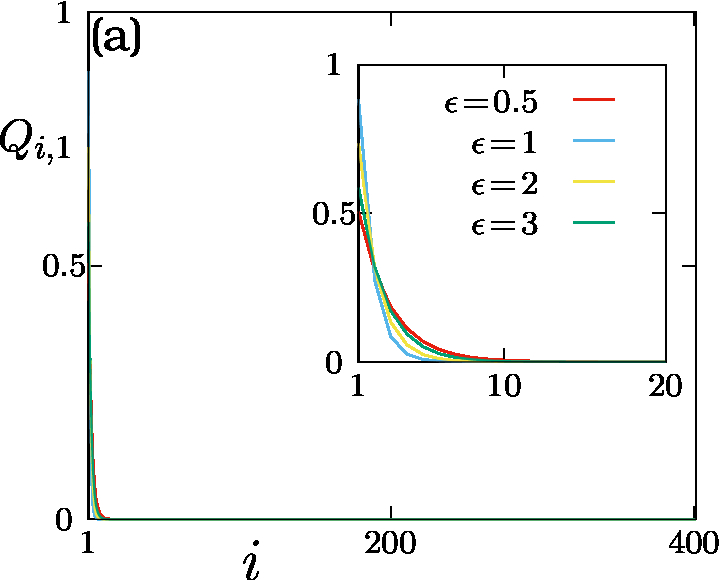}
\includegraphics[height=0.4\textwidth,width=0.41\textwidth]{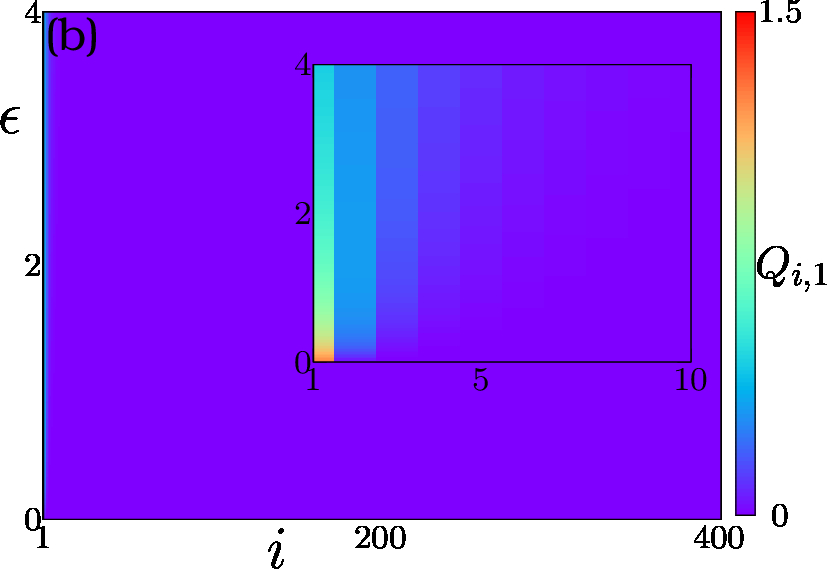}
\end{center}
\caption{Response amplitudes of a network of $n = 400$ coupled plasma oscillators. Only the first 20 oscillators are shown; beyond approximately the 10th oscillator, oscillations vanish up to the 400th unit. (a) Response amplitude as a function of oscillator index for different values of $\epsilon$. (b) Response amplitude as a function of oscillator index and coupling strength $\epsilon$.}
\label{fig2}
\end{figure}

Following the procedure described in the previous section, we extend the analysis to a network of $n=400$ oscillators and compute the corresponding response amplitudes. These amplitudes are plotted to illustrate the system behavior, as shown in Fig.~\ref{fig2}.

In Fig.~\ref{fig2}(a), it is observed that the transmission of the weak periodic signal decays rapidly and becomes negligible beyond approximately the first ten oscillators, indicating strongly attenuated propagation across the network. Similarly, Fig.~\ref{fig2}(b) shows that variations in the coupling strength have only a limited influence on the response amplitudes, suggesting that uniform coupling does not significantly enhance signal transmission in this configuration.

To further investigate signal transmission in bidirectionally coupled oscillators, we extend the analysis by introducing asymmetric coupling strengths between neighboring units. The impact of such asymmetry on the response amplitudes and transmission efficiency is examined in the following section.

\section{Plasma Oscillator Networks with Asymmetric Coupling}
\subsection{Two Coupled Plasma Oscillators with Asymmetric Coupling}

We consider two coupled plasma oscillators with distinct coupling strengths $\epsilon_1$ and $\epsilon_2$ \cite{karunanidhi2025asymmetric}. In contrast to the symmetric case, the interaction is nonreciprocal, characterized by asymmetric coupling coefficients. As in the previous section, only the first oscillator is driven by a weak external periodic forcing. The corresponding equations of motion are
\begin{eqnarray}
\ddot{x}_1 + d(1+x_1^2)\dot{x}_1 + \omega_0^2 x_1 + \alpha x_1^2 + \beta x_1^3 &=& F \cos(\omega t) + \epsilon_2 (x_2 - x_1), \nonumber \\
\ddot{x}_2 + d(1+x_2^2)\dot{x}_2 + \omega_0^2 x_2 + \alpha x_2^2 + \beta x_2^3 &=& \epsilon_1 (x_1 - x_2).
\label{eq9}
\end{eqnarray}

Before proceeding, we note that each unit corresponds to a modified Duffing oscillator with position-dependent damping and an asymmetric potential characterized by the parameter $\alpha$. Depending on the values and signs of $\alpha$, $\beta$, and $\omega_0^2$, the uncoupled potential may exhibit either single-well or double-well configurations, with symmetric or asymmetric profiles.

To illustrate the nonreciprocal nature of the coupling, we consider the reduced static system obtained by neglecting the inertial and damping terms and omitting the external forcing. Equation~\eqref{eq9} then reduces to
\begin{eqnarray}
\omega_0^2 x_1 + \alpha x_1^2 + \beta x_1^3 + \epsilon_2 x_1 &=& \epsilon_2 x_2, \nonumber \\
\omega_0^2 x_2 + \alpha x_2^2 + \beta x_2^3 + \epsilon_1 x_2 &=& \epsilon_1 x_1.
\end{eqnarray}

Defining
\begin{eqnarray}
F_1(x_1,x_2) &=& \omega_{\epsilon_2} x_1 + \alpha x_1^2 + \beta x_1^3 - \epsilon_2 x_2, \nonumber \\
F_2(x_1,x_2) &=& \omega_{\epsilon_1} x_2 + \alpha x_2^2 + \beta x_2^3 - \epsilon_1 x_1,
\end{eqnarray}
with $\omega_{\epsilon_1}=\omega_0^2+\epsilon_1$ and $\omega_{\epsilon_2}=\omega_0^2+\epsilon_2$, we obtain
\begin{equation}
\frac{\partial F_1}{\partial x_2} = -\epsilon_2, 
\qquad
\frac{\partial F_2}{\partial x_1} = -\epsilon_1,
\end{equation}
and therefore
\begin{equation}
\frac{\partial F_1}{\partial x_2} \neq \frac{\partial F_2}{\partial x_1},
\label{eq10}
\end{equation}
which confirms the nonreciprocal character of the coupling.

If $\epsilon=\epsilon_1=\epsilon_2$, then $\omega_{\epsilon_1}=\omega_{\epsilon_2}=\omega_\epsilon=\omega_0^2+\epsilon$, and the system admits a scalar potential of the form
\begin{equation}
V(x_1,x_2)=\frac{\omega_\epsilon}{2}(x_1^2+x_2^2)
+\frac{\alpha}{3}(x_1^3+x_2^3)
+\frac{\beta}{4}(x_1^4+x_2^4)
-\epsilon x_1x_2.
\label{eq11}
\end{equation}

\begin{figure}[h]
\begin{center}
\includegraphics[width=0.4\textwidth]{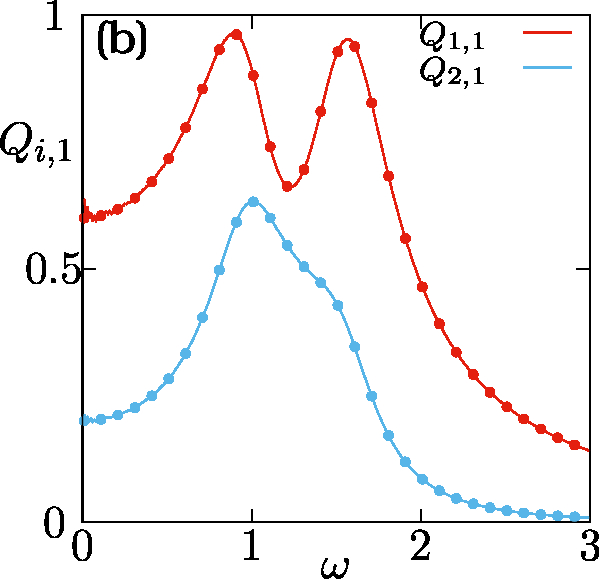}
\end{center}
\caption{Numerically (solid lines) and analytically (dots) computed response amplitudes of the plasma oscillators for $\epsilon_1=0.5$ and $\epsilon_2=1$.}
\label{fig3}
\end{figure}

In the present case, however, $\epsilon_1\neq\epsilon_2$. This asymmetry [cf. Eq.~\eqref{eq10}] implies nonreciprocal interaction, and therefore the coupled system is nonconservative. In what follows, we exploit this nonreciprocity as a mechanism for signal transmission between oscillators. We focus on the response at the driving frequency $\omega$ and assume solutions of the form
\begin{eqnarray}
x_1 &=& A_1\cos(\omega t+\phi_1), \nonumber\\
x_2 &=& A_2\cos(\omega t+\phi_2).
\label{eq12}
\end{eqnarray}
Here, $A_i$ and $\phi_i$ ($i=1,2$) denote the amplitude and phase of the steady-state response at frequency $\omega$. Substituting the ansatz~\eqref{eq12} into Eq.~\eqref{eq9} and applying the method of harmonic balance \cite{rajasekar2016nonlinear, rajamani2013signal}, we obtain the following coupled nonlinear algebraic equations for $A_1$, $A_2$, $\phi_1$, and $\phi_2$:
\begin{eqnarray}
A_1^2\left[u_1^2+\left(1+\frac{A_1^2}{2}\right)^2 d^2\omega^2\right]
+ A_2^2 \epsilon_2^2
- 2A_2^2\left(u_1 u_2 - \left(1+\frac{A_1^2}{2}\right)^2 d^2\omega^2\right)
&=& F^2, \nonumber\\
A_2^2\left[u_2^2+\left(1+\frac{A_2^2}{2}\right)^2 d^2\omega^2\right]
- \epsilon_1^2 A_1^2 &=& 0,
\label{eq13}
\end{eqnarray}
with
\begin{eqnarray}
\phi_1 &=& \arctan\left[\frac{\left(1+\frac{A_1^2}{2}\right)(\epsilon_1 A_1^2+\epsilon_2 A_2^2)d\omega}{A_1\epsilon_1 u_1 - A_2\epsilon_2 u_2}\right], \nonumber\\
\phi_2 &=& \phi_1 - \arctan\left[-\,\frac{\left(1+\frac{A_2^2}{2}\right)d\omega}{u_2}\right],
\label{eq14}
\end{eqnarray}
where
\begin{eqnarray}
u_1 &=& \omega_0^2 - \omega^2 + \epsilon_1 + \frac{3}{4}\beta A_1^2, \nonumber\\
u_2 &=& \omega_0^2 - \omega^2 + \epsilon_2 + \frac{3}{4}\beta A_2^2.
\label{eq15}
\end{eqnarray}

The response amplitudes of both oscillators, obtained numerically and analytically under asymmetric coupling, are shown in Fig.~\ref{fig3}. Both oscillators exhibit a nonlinear resonance response. In Fig.~\ref{fig3}, the red and blue curves correspond to oscillators~1 and~2, respectively, for $\epsilon_1=0.5$ and $\epsilon_2=1.0$. In this regime, the response of oscillator~1 to the weak periodic forcing is higher than that of oscillator~2. Consequently, for the chosen parameter values, the amplitude $Q_1$ exceeds $Q_2$.

\subsection{Network of Bidirectionally Coupled Oscillators with Asymmetric Coupling}

\begin{figure}[h]
\centering
\includegraphics[width=0.4\textwidth]{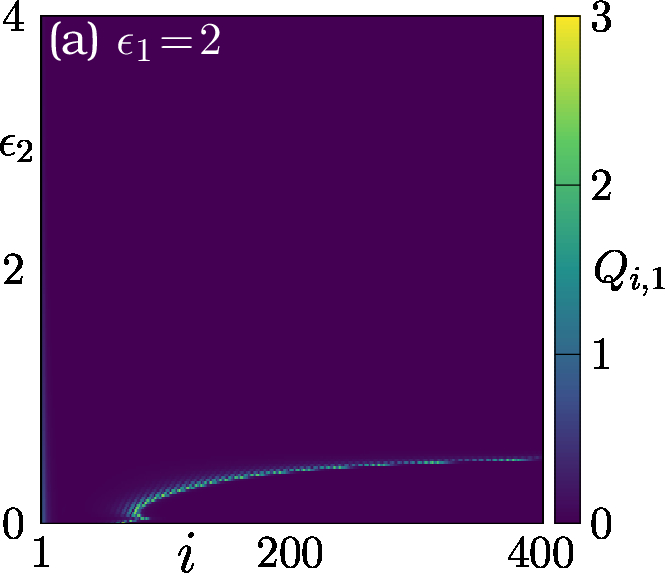}
\includegraphics[width=0.4\textwidth]{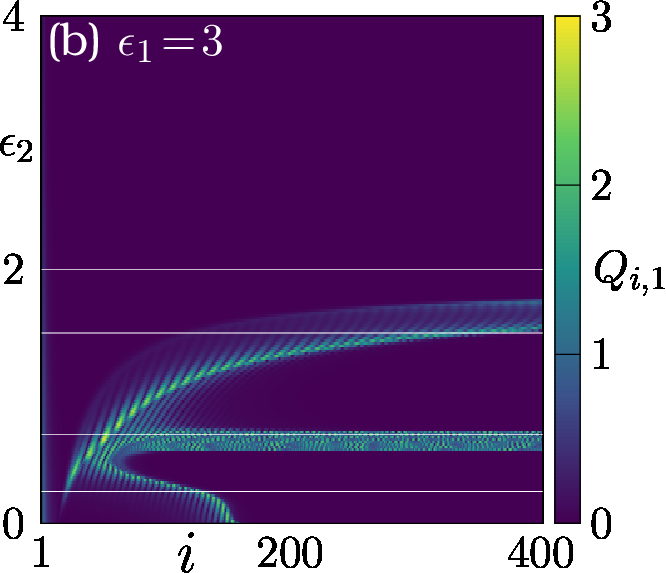}
\includegraphics[width=0.4\textwidth]{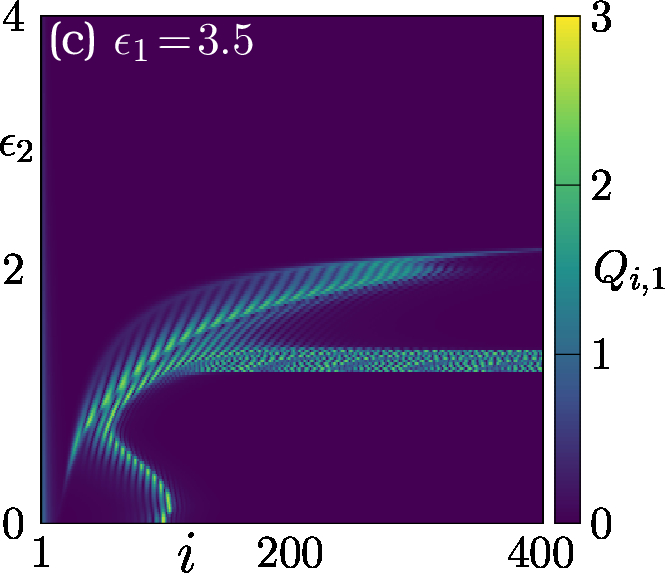}
\includegraphics[width=0.4\textwidth]{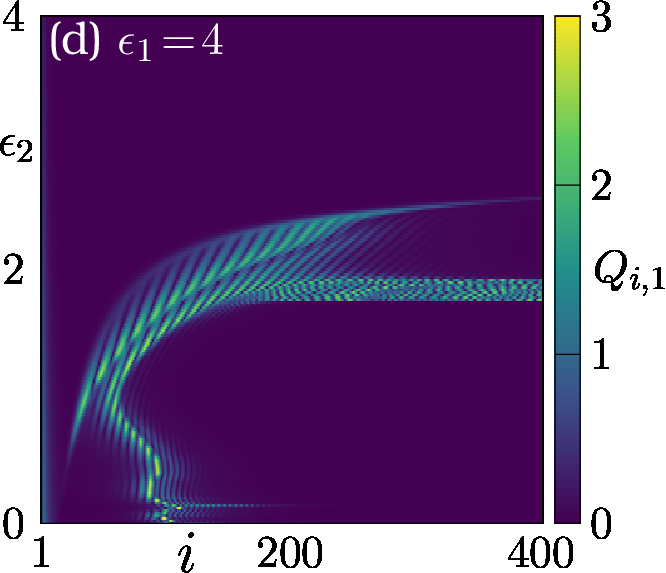}
\caption{Surface plots of the response amplitude $Q_i$ at the driving frequency $\omega$ as a function of oscillator index $i$ and backward coupling strength $\epsilon_2$, for fixed forward coupling strengths $\epsilon_1$: (a) $2$, (b) $3$, (c) $3.5$, and (d) $4$. The red dashed horizontal lines in panel (c) indicate the selected values $\epsilon_2 = 0.25,\,0.7,\,1.5,$ and $2$.}
\label{fig4}
\end{figure}

We extend the analysis from two coupled plasma oscillators to a network of oscillators with asymmetric interactions, characterized by a forward (feedforward) coupling strength $\epsilon_1$ and a backward (feedback) coupling strength $\epsilon_2$. In this configuration, the coupling is nonreciprocal, and the interaction between oscillators is direction-dependent. The governing equations of motion are given by
\begin{eqnarray}
\ddot{x}_1 + d(1 + x_1^2)\dot{x}_1 + \omega_0^2 x_1 + \alpha x_1^2 + \beta x_1^3 &=& F \cos(\omega t) + \epsilon_2 (x_2 - x_1), \nonumber \\
\ddot{x}_i + d(1 + x_i^2)\dot{x}_i + \omega_0^2 x_i + \alpha x_i^2 + \beta x_i^3 &=& \epsilon_1 (x_{i-1} - x_i) + \epsilon_2 (x_{i+1} - x_i), \nonumber \\
\ddot{x}_n + d(1 + x_n^2)\dot{x}_n + \omega_0^2 x_n + \alpha x_n^2 + \beta x_n^3 &=& \epsilon_1 (x_{n-1} - x_n),
\label{eq16}
\end{eqnarray}
for $i = 2,3,\ldots,n-1$.

To investigate signal transmission across the network, the response amplitude $Q_i$, defined in Eq.~\eqref{eq6}, is evaluated. Figure~\ref{fig4} shows surface plots of $Q_i$ at the driving frequency $\omega$ for a network of size $n=400$. The forward coupling strength is fixed at $\epsilon_1 = 2,\,3,\,3.5,$ and $4$, while the backward coupling strength $\epsilon_2$ is varied.

\textbf{Transmission regimes and their classification.}
To facilitate interpretation of Fig.~\ref{fig4}, we classify the signal transmission into three distinct regimes:

\begin{itemize}
\item \textbf{Type-I (localized transmission):} The signal propagates only over a limited subset of oscillators and decays thereafter.
\item \textbf{Type-II (efficient transmission):} The signal propagates across the entire network with negligible attenuation.
\item \textbf{Type-III (no transmission):} The signal is rapidly attenuated near the boundary and does not propagate through the network.
\end{itemize}

\textbf{Dependence on coupling strengths.}
For $\epsilon_1 = 2$ [Fig.~\ref{fig4}(a)], nonzero values of $Q_i$ appear only within a narrow range of $\epsilon_2$. When $\epsilon_2 < 0.55$, the system exhibits type-I behavior, with strongly attenuated and localized transmission.

For $\epsilon_1 = 3$ [Fig.~\ref{fig4}(b)], all three regimes are clearly observed. Specifically:
\begin{itemize}
\item $\epsilon_2 < 0.57$: type-I (localized transmission),
\item $\epsilon_2 \in [0.57,\,0.73]$: type-II (efficient transmission),
\item $0.73 < \epsilon_2 < 1.77$: type-I (re-entrant localized regime),
\item $\epsilon_2 > 1.77$: type-III (no transmission).
\end{itemize}

For higher values of $\epsilon_1$, the same three regimes persist, with a shift in the onset of the no-transmission regime. In particular, the transition to type-III occurs at $\epsilon_2 \approx 1.77$ for $\epsilon_1 = 3$, $\epsilon_2 \approx 2.17$ for $\epsilon_1 = 3.5$ [Fig.~\ref{fig4}(c)], and $\epsilon_2 \approx 2.56$ for $\epsilon_1 = 4$ [Fig.~\ref{fig4}(d)].

A general trend observed in Fig.~\ref{fig4} is that effective transmission (type-I and type-II) requires $\epsilon_1 > \epsilon_2$. When $\epsilon_2 \ge \epsilon_1$, the system predominantly exhibits type-III behavior, with no signal propagation.

In both type-I and type-II regimes, a characteristic initial attenuation is observed over the first few oscillators, followed by a gradual growth of the signal amplitude along the chain. This appears in Fig.~\ref{fig4} as a low-amplitude region near the origin, indicating delayed signal emergence.

\textbf{Representative cases.}
To examine these regimes in detail, we select four representative values of $\epsilon_2$ from Fig.~\ref{fig4}(b): $\epsilon_2 = 0.25$ (type-I), $0.7$ (type-II), $1.5$ (type-I), and $2$ (type-III), indicated by red dashed horizontal lines.

\begin{figure}[h]
\begin{center}
\includegraphics[width=0.45\textwidth]{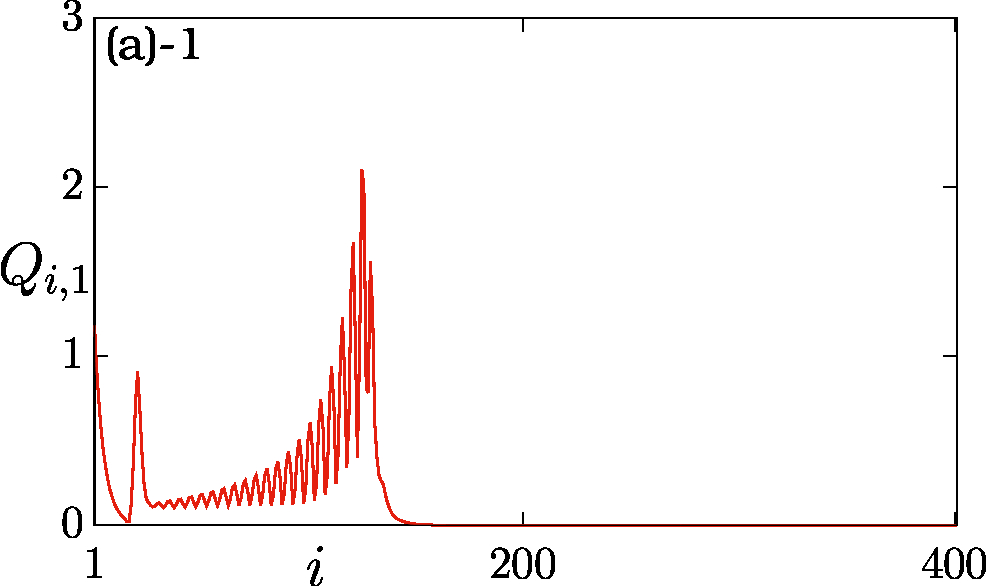}
\includegraphics[width=0.45\textwidth]{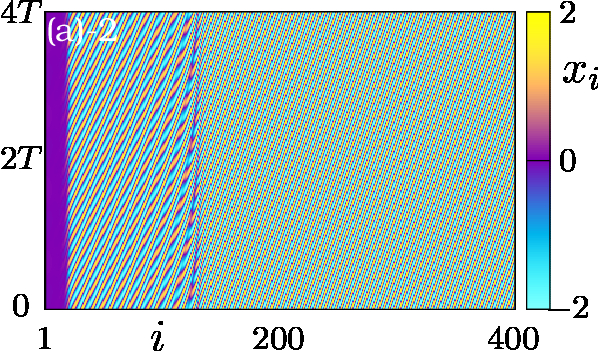}
\includegraphics[width=0.45\textwidth]{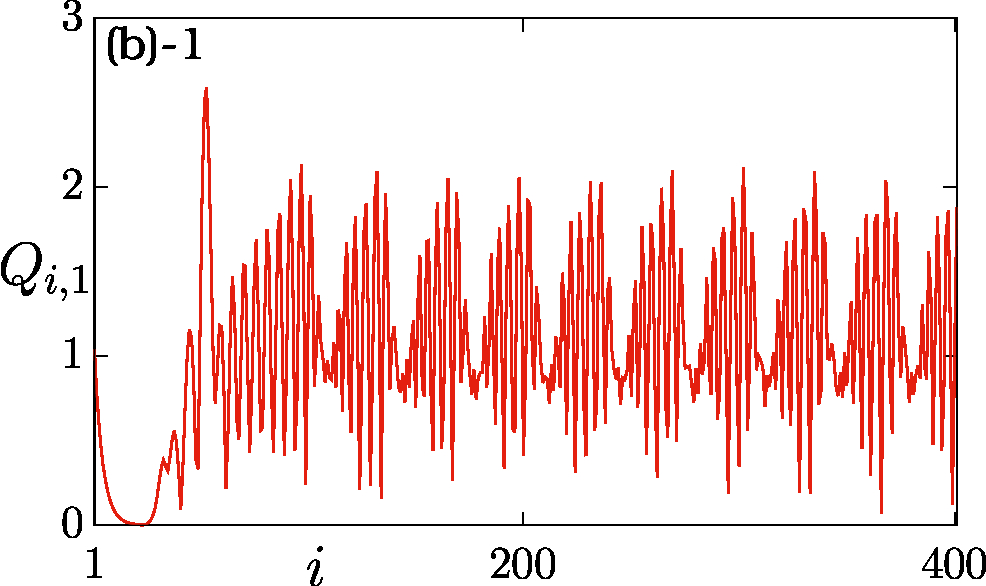}
\includegraphics[width=0.45\textwidth]{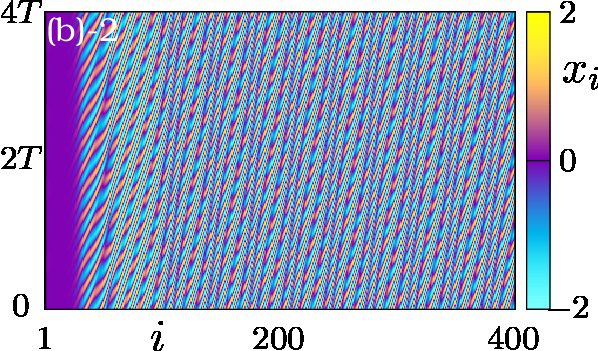}
\includegraphics[width=0.45\textwidth]{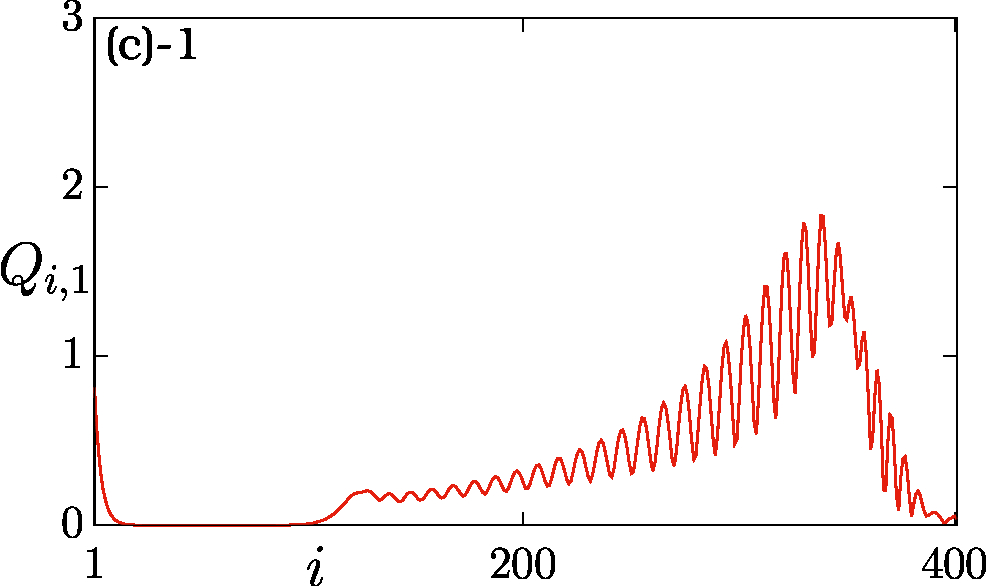}
\includegraphics[width=0.45\textwidth]{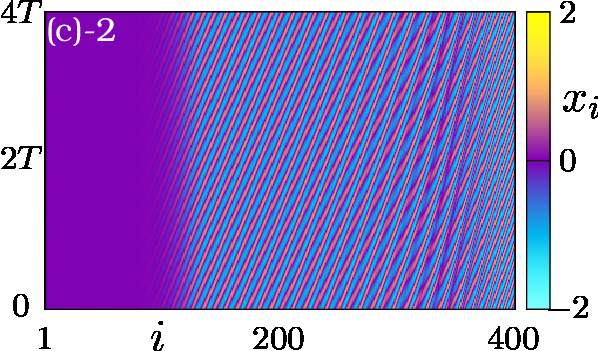}
\includegraphics[width=0.45\textwidth]{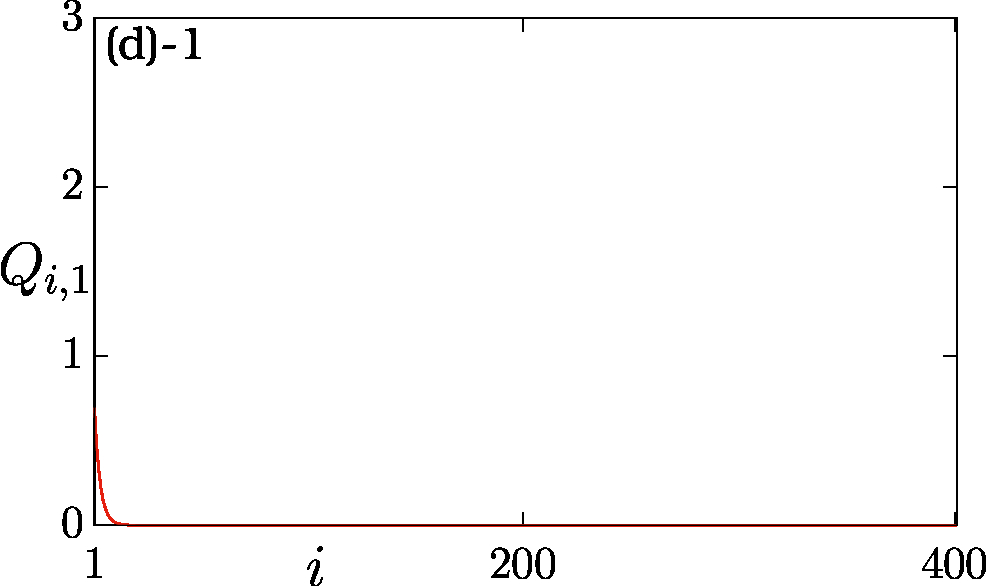}
\includegraphics[width=0.45\textwidth]{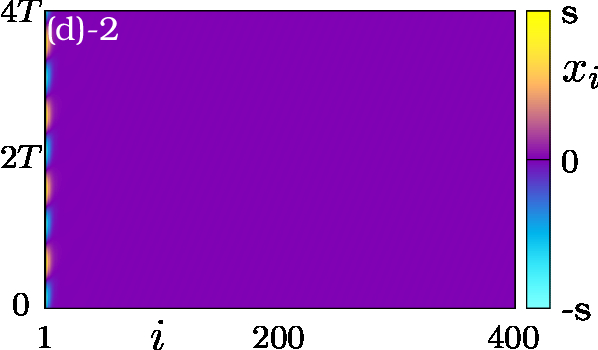}
\end{center}
\caption{Spatial and spatiotemporal dynamics of signal transmission for $\epsilon_1=3$ with $\epsilon_2=0.25,\,0.7,\,1.5,$ and $2$. Panels (a--d) show the response amplitude $Q_i$ as a function of oscillator index $i$, illustrating type-I [(a),(c)], type-II [(b)], and type-III [(d)] transmission. Panels (e--h) show the corresponding spatiotemporal evolution of $x_i$. Non-uniform amplitude distributions and modulated wave patterns in (b,c) indicate the presence of higher harmonics, while (d) shows rapid attenuation near the boundary.}
\label{fig5}
\end{figure}

\textbf{Spatial dynamics.}
Figure~\ref{fig5}(a--d) shows the spatial variation of the response amplitude $Q_i$ (see Eq.~\eqref{eq6}) for $\epsilon_1=3$ and the selected values of $\epsilon_2$.

For $\epsilon_2 = 0.25$ [Fig.~\ref{fig5}(a)], the system exhibits type-I transmission: the signal grows over a limited range of oscillators and then decays, indicating localized propagation. A similar type-I regime is observed for $\epsilon_2 = 1.5$ [Fig.~\ref{fig5}(c)], although the signal re-emerges after an initial decay and propagates with a non-uniform amplitude distribution, suggesting the presence of higher harmonics.

For $\epsilon_2 = 0.7$ [Fig.~\ref{fig5}(b)], the system displays type-II transmission, characterized by sustained propagation across the network. In contrast, for $\epsilon_2 = 2$ [Fig.~\ref{fig5}(d)], type-III behavior is observed, with rapid attenuation near the boundary.

\textbf{Spatiotemporal dynamics.}
The corresponding spatiotemporal evolution, shown in Fig.~\ref{fig5}(e--h), supports these observations. In Fig.~\ref{fig5}(e), although the primary signal attenuates beyond a certain index, residual oscillations persist, indicating higher harmonic contributions. Similar complex wave patterns appear in Fig.~\ref{fig5}(f,g), reflecting nonlinear interactions and harmonic generation. In contrast, Fig.~\ref{fig5}(h) confirms the absence of transmission, with oscillations confined near the boundary.

These results highlight the role of asymmetric coupling in enabling diverse transmission regimes and promoting nonlinear wave interactions. In the next section, we analyze harmonic generation in more detail.

\section{Higher Harmonic Generation and Reversal of Vibrational Resonance}

\begin{figure}[h]
\begin{center}
\includegraphics[width=0.33\textwidth,height=0.35\textwidth]{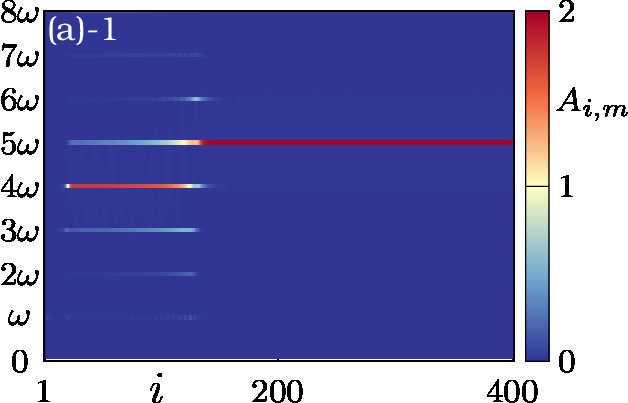} 
\includegraphics[width=0.33\textwidth,height=0.35\textwidth]{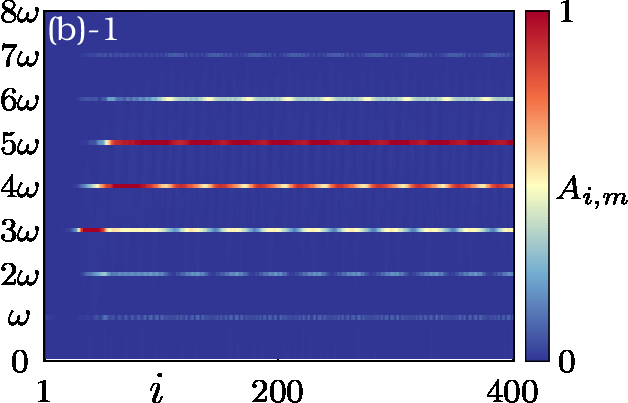}
\includegraphics[width=0.33\textwidth,height=0.35\textwidth]{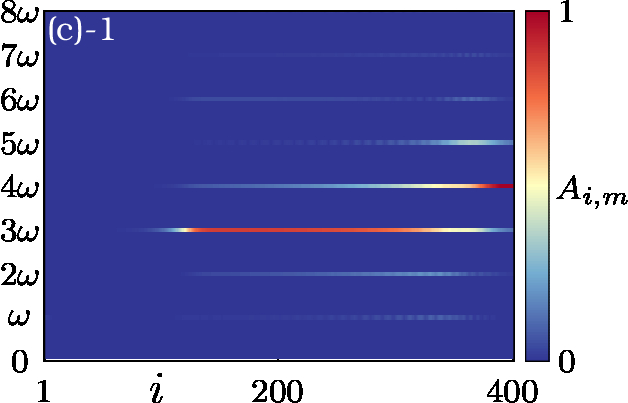}
\end{center}
\caption{Color maps of the mode amplitude $A_i$ as a function of oscillator index $i$ and frequency $\omega$ for $\epsilon_1 = 3$ and different values of $\epsilon_2$: (a) $0.25$, (b) $0.7$, (c) $1.5$, and (d) $2$. Panels (a) and (c) correspond to type-I transmission, panel (b) to type-II transmission, and panel (d) to type-III (no transmission). The color bar indicates the magnitude of $A_i$.}
\label{fig6}
\end{figure}

In this section, we analyze the emergence of higher harmonics in the transmitted signal along a network of bidirectionally coupled plasma oscillators. The parameters are the same as in the previous section, with asymmetric coupling strengths $\epsilon_1 = 3$ and $\epsilon_2 = 0.25,\,0.7,\,1.5,$ and $2$. To characterize the spectral content, we compute the amplitude spectrum $A_i$ for each oscillator and represent it as a function of the oscillator index $i$ and frequency.

\begin{figure}[h]
\begin{center}
\includegraphics[width=0.515\textwidth]{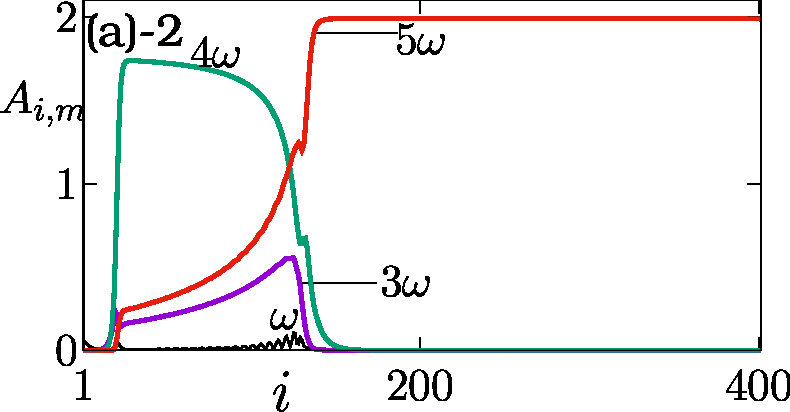}
\includegraphics[width=0.515\textwidth]{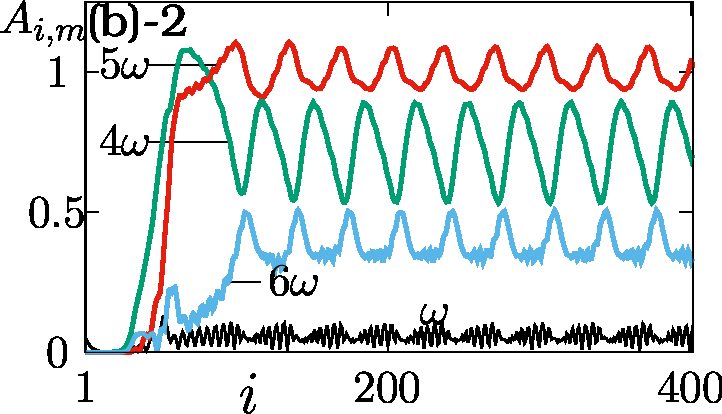}
\includegraphics[width=0.505\textwidth]{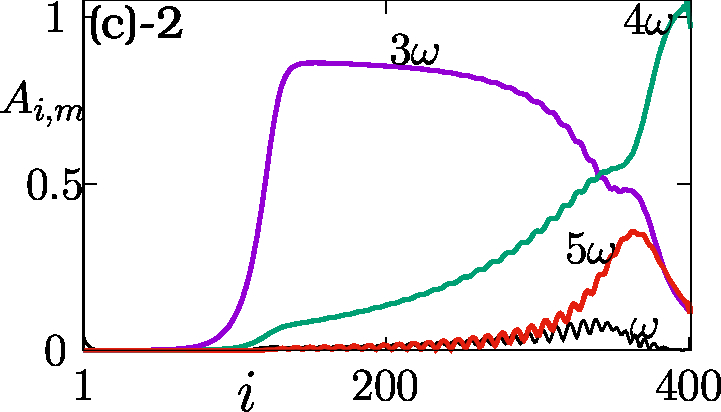}
\end{center}
\caption{Amplitude spectrum $A_i$ at selected harmonic frequencies for $\epsilon_1 = 3$ and (a) $\epsilon_2 = 0.25$, (b) $\epsilon_2 = 0.7$, and (c) $\epsilon_2 = 1.5$.}
\label{fig7}
\end{figure}

For $\epsilon_2 = 0.25$ [Fig.~\ref{fig6}(a)], the fundamental frequency $\omega$ decays rapidly and becomes negligible beyond approximately the $157$th oscillator (type-I transmission). A similar attenuation is observed for lower-order harmonics such as $3\omega$ and $4\omega$. In contrast, the $5\omega$ component exhibits a qualitatively different behavior: it grows in amplitude and propagates across the entire network without decay. Beyond $i \approx 157$, its amplitude saturates at an approximately constant value ($A \approx 1.98$), indicating robust transmission. This feature is further illustrated in Fig.~\ref{fig7}(a), where the dominance of the $5\omega$ harmonic is evident.
This behavior can be interpreted within the framework of vibrational resonance (VR) \cite{landa2000vibrational}. In conventional VR, a weak low-frequency signal is amplified through interaction with an externally applied high-frequency drive. In the present system, however, no such external forcing is present. Instead, the asymmetric coupling induces intrinsic nonlinear interactions that generate higher harmonics. Beyond a threshold index ($i \approx 157$), the fundamental mode is suppressed, while a higher harmonic (notably $5\omega$) becomes dominant and propagates with nearly constant amplitude. This constitutes a \emph{reversal of vibrational resonance}, in which energy is transferred from the fundamental mode to higher harmonics, leading to selective amplification. The observed amplification, reaching up to $\sim 40$ times the input amplitude, highlights the efficiency of this nonlinear transfer mechanism.

For $\epsilon_2 = 0.7$ [Fig.~\ref{fig6}(b)], multiple harmonics, including the fundamental frequency $\omega$, are sustained throughout the network (type-II transmission). As shown in Fig.~\ref{fig7}(b), higher harmonics such as $4\omega$ and $5\omega$ persist along the chain with modulated amplitudes, indicating strong nonlinear mixing.

For $\epsilon_2 = 1.5$ [Fig.~\ref{fig6}(c)], selective transmission is observed (type-I transmission), where only certain harmonics survive. As illustrated in Fig.~\ref{fig7}(c), the $3\omega$ component initially dominates but gradually decays, while higher harmonics such as $4\omega$ and $5\omega$ emerge and grow toward the end of the network, indicating a redistribution of energy among modes.

For $\epsilon_2 = 2$ [Fig.~\ref{fig6}(d)], no significant spectral components are observed beyond the initial oscillators, confirming the absence of signal transmission (type-III regime).


\section{Transmission Phase Diagram under Asymmetric Coupling}

\begin{figure}[h]
\centering
\includegraphics[width=0.4\textwidth]{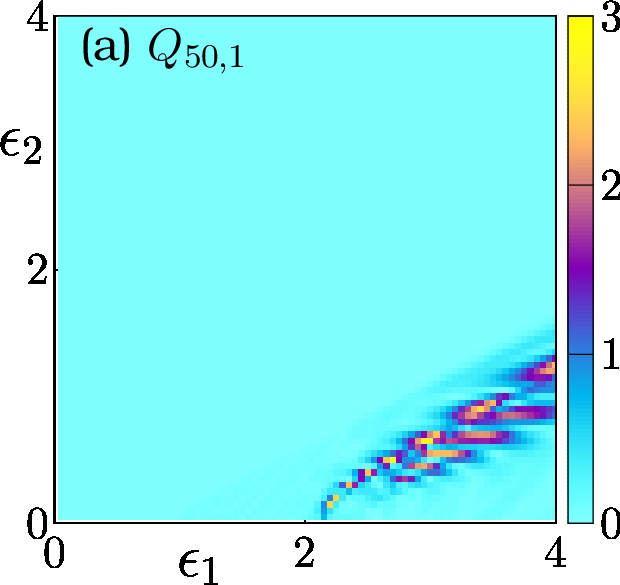}
\includegraphics[width=0.4\textwidth]{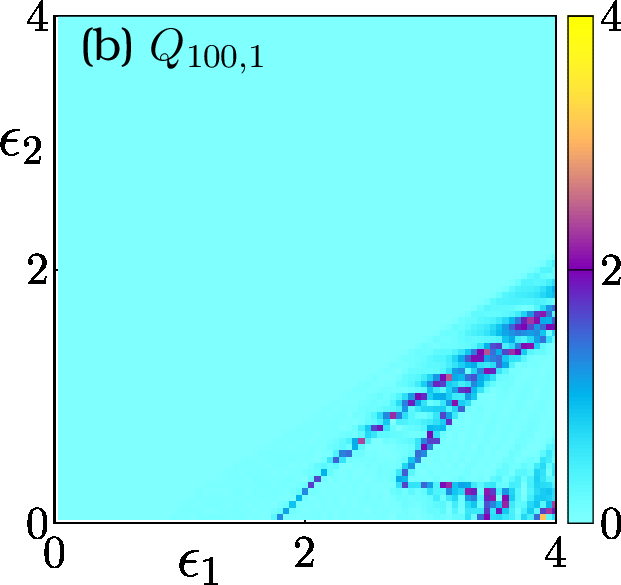}
\includegraphics[width=0.4\textwidth]{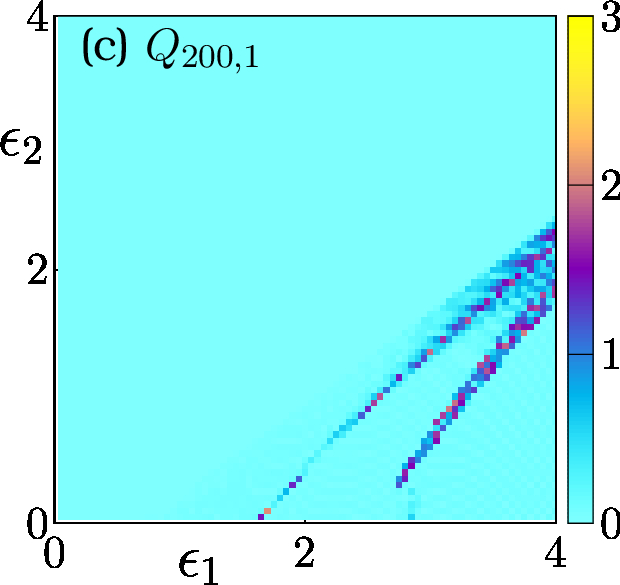}
\includegraphics[width=0.4\textwidth]{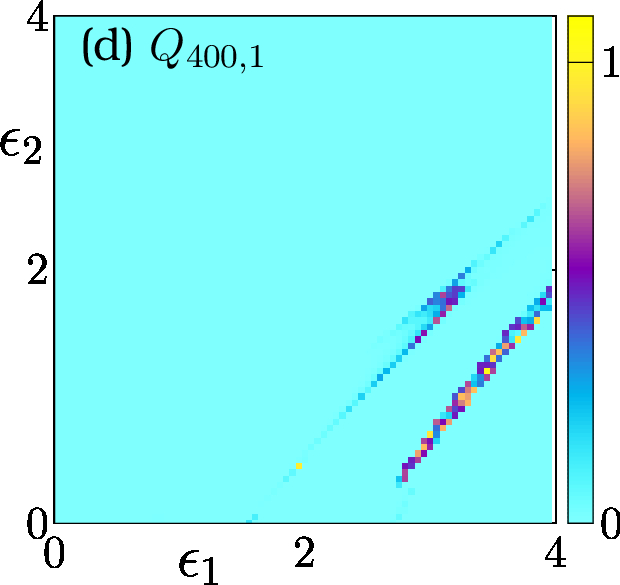}
\caption{Two-parameter response maps of the amplitude $Q_i$ for selected oscillators: (a) $i = 50$, (b) $i = 100$, (c) $i = 150$, and (d) $i = 200$, as functions of $\epsilon_1$ and $\epsilon_2$. Regions of high amplitude occur predominantly at lower values of $\epsilon_2$, indicating weakly damped regimes. As the oscillator index increases, these regions become progressively narrower, identifying the parameter ranges that support long-range signal transmission.}
\label{fig8}
\end{figure}

To further quantify the role of asymmetric coupling in signal propagation, we examine two-parameter response maps of the amplitude $Q_i$ for selected oscillators at different positions along the chain. Figure~\ref{fig8} shows the variation of $Q_i$ in the $(\epsilon_1,\epsilon_2)$ parameter space for $i = 50,\,100,\,150,$ and $200$.

For smaller oscillator indices [Fig.~\ref{fig8}(a,b)], the response amplitude occupies a relatively broad region of parameter space, with significant values observed even for moderate $\epsilon_2$. As the oscillator index increases [Fig.~\ref{fig8}(c,d)], the regions of high amplitude become progressively narrower and shift toward lower values of $\epsilon_2$. This behavior indicates that sustained signal propagation over longer distances requires sufficiently strong forward coupling ($\epsilon_1$) combined with relatively weak backward coupling ($\epsilon_2$).

In particular, for larger $i$, nonzero responses are confined to a thin band in the $(\epsilon_1,\epsilon_2)$ plane, highlighting the increasingly restrictive conditions for long-range transmission. Outside these regions, the signal is rapidly attenuated, consistent with the type-III transmission regime identified earlier. These results confirm that asymmetric coupling is a key mechanism for enabling efficient energy transport and selective signal propagation across the network.

\begin{figure}[h]
\centering
\includegraphics[width=0.35\textwidth]{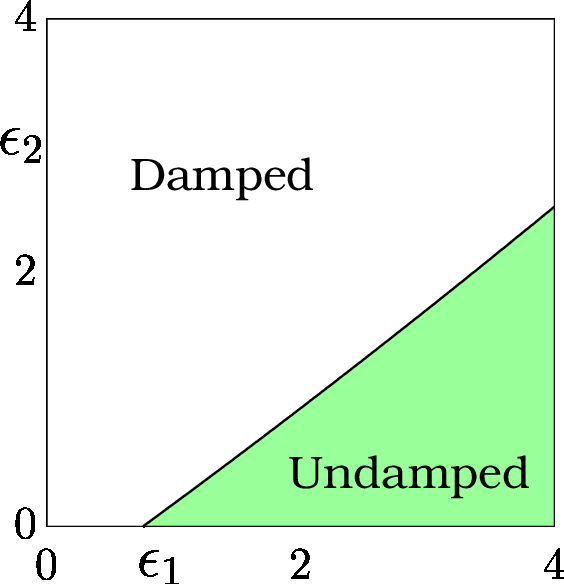}
\caption{Phase diagram in the $(\epsilon_1,\epsilon_2)$ plane showing the boundary between damped and undamped transmission regimes.}
\label{fig9}
\end{figure}

The phase diagram in the $(\epsilon_1,\epsilon_2)$ parameter space, shown in Fig.~\ref{fig9}, delineates the boundary between damped and undamped transmission regimes. The undamped region corresponds to parameter combinations for which the signal propagates across the network with negligible attenuation, either through the fundamental frequency or via higher harmonic components generated by nonlinear interactions. In contrast, the damped region corresponds to conditions under which the signal is rapidly attenuated within the first few oscillators.

As indicated by the phase boundary, sustained transmission occurs predominantly when $\epsilon_1 > \epsilon_2$, underscoring the importance of coupling asymmetry for long-range energy transport. Notably, the undamped regime also includes regions where higher harmonics dominate the transmission, reflecting the redistribution of energy from the fundamental mode to higher-frequency components. Overall, the phase diagram provides a comprehensive characterization of the parameter regimes that support efficient signal propagation and harmonic selection.

\section{Conclusion}

In this work, we have investigated signal transmission in a one-dimensional chain of $n$ locally coupled plasma oscillators with asymmetric bidirectional coupling characterized by $(\epsilon_1,\epsilon_2)$. The results show that symmetric coupling suppresses signal propagation, whereas coupling asymmetry enables efficient nonlinear transmission across the network. In particular, sustained transmission is observed when the forward coupling strength $\epsilon_1$ exceeds the backward coupling strength $\epsilon_2$, as confirmed by the phase diagram in the $(\epsilon_1,\epsilon_2)$ parameter space.

A central finding is the intrinsic generation of higher harmonics through nonlinear interactions, accompanied by a reversal of vibrational resonance, in which energy is transferred from the fundamental mode to higher-order harmonics. The interplay between coupling asymmetry and nonlinearity determines both the onset of transmission and the selection of the dominant spectral components.

These results provide insight into mechanisms for controlling signal propagation and harmonic amplification in nonlinear oscillator networks, with potential relevance for tunable energy transport and frequency-selective devices.

\section*{Acknowledgments}

This work has received financial support by the Spanish State Research Agency (AEI) and the European Regional Development Fund (ERDF, EU) under Project No. PID2023-148160NB-I00 (MCIN/AEI/10.13039/501100011033).



\end{document}